\documentclass[11pt, a4paper]{article}
\usepackage{epsfig}

\newcommand{\HH}{{\cal H}}

\def \aa{\hat{a}}

\begin{document}

\title{Indistinguishable Particles in Quantum Mechanics: An Introduction}

\author{Yasser Omar\\
\textit{Departamento de Matem\'{a}tica}\\\textit{ISEG,
Universidade T\'{e}cnica de Lisboa}\\\textit{P-1200-781 Lisbon,
Portugal}\\\textit{\small{Yasser.Omar@iseg.utl.pt}}\\}

\date{15 May 2005}

\maketitle

\begin{abstract}
In this article, we discuss the identity and indistinguishability
of quantum systems and the consequent need to introduce an extra
postulate in Quantum Mechanics to correctly describe situations
involving indistinguishable particles. This is, for electrons, the
Pauli Exclusion Principle, or in general, the Symmetrization
Postulate. Then, we introduce fermions and bosons and the
distributions respectively describing their statistical behaviour
in indistinguishable situations. Following that, we discuss the
spin-statistics connection, as well as alternative statistics and
experimental evidence for all these results, including the use of
bunching and antibunching of particles emerging from a beam
splitter as a signature for some bosonic or fermionic states.
\end{abstract}

\section{An Extra Postulate is Required}
\label{Sec. Pauli}

I believe most physicists would consider that the postulates (or
at least the properties they embody) concerning the superposition,
evolution and measurement of quantum states cover the essence of
Quantum Mechanics, the theory that is at the basis of current
fundamental Physics and gives us such an accurate description of
Nature at the atomic scale. Yet, if the theory was only based on
these postulates (or properties), its descriptive power would be
almost zero and its interest, if any, would be mainly
mathematical. As soon as one wants to describe matter, one has to
include an extra postulate: \emph{Pauli's Exclusion Principle}.
One of its usual formulations, equivalent to the one proposed
originally by Wolfang Pauli in 1925
\cite{pauli-exclusion}, is the following:\\

\textbf{Pauli's Exclusion Principle} --- \emph{No two electrons
can share the same quantum numbers.}\\

This principle refers to electrons, which constitute a significant
(but not the whole) part of matter, and is crucial in helping us
explain a wide range of phenomena, including:

\begin{itemize}

\item The electronic structure of atoms and, as a consequence, the
whole Periodic Table;

\item The electronic structure of solids and their electrical and
thermal pro\-perties;

\item The formation of white dwarfs, where the gravitational
collapse of the star is halted by the pressure resulting from its
electrons being unable to occupy the same states;

\item The repulsive force that is part of the \emph{ionic bond} of
molecules and puts a limit to how close the ions can get (e.g.,
$0.28\!$ nm between $Na^+$ and $Cl^-$ for solid sodium chloride),
given the restrictions to the states the overlapping electrons can
share.

\end{itemize}

We thus see how Pauli's insight when proposing the Exclusion
Principle was fundamental for the success of Quantum Mechanics.
Although he made many other important contributions to Physics, it
was for this one that he was awarded the Nobel prize in 1945.

Historically, it is also interesting to note that this happened
before Samuel Goudsmit and Georg Uhlenbeck introduced the idea of
\emph{electron spin} \cite{goudsmit, uhlenbeck} later in 1925. In
1913, Niels Bohr presented his model for the electronic structure
of the atom to explain the observed discrete energy spectra of
Hydrogen and other elements \cite{bohr-model}: the electrons fly
around the positive nucleus in circular orbits\footnote{A wrong
idea, as we shall discuss later.} with quantized angular momentum.
This quantization restricts the possible orbits to a discrete set,
each corresponding to an energy level of the atom. This model was
then improved during the following decade, mainly by Arnold
Sommerfeld and Alfred Land\'e, rendering it more sophisticated,
trying to make it able to account for the multiplet structure of
spectral lines, including for atoms in electric and magnetic
fields. In 1922, Pauli joins the effort (actually, it was rather a
competition) to find an explanation for the then-called
\emph{anomalous Zeeman effect}, a splitting of spectral lines of
an atom in a magnetic field that was different from the already
known Zeeman splitting (see \cite{tomonaga} for a technical
historical account on this competition). Another puzzle at the
time, identified by Bohr himself, was the following: how to
explain that in an atom in the ground state the electrons do not
all populate the orbit closest to the nucleus (corresponding to
the lowest energy level) \cite{bohr}? These two problems led Pauli
to postulate, towards the end of 1924, a new property for the
electron
--- \emph{``a two-valuedness not describable classically"}
\cite{pauli-pre} --- and soon after the Exclusion Principle
\cite{pauli-exclusion} as fundamental rules for the classification
of spectral lines. But Pauli did not present any model for this
extra degree of freedom of the electrons. A few months later,
Goudsmit and Uhlenbeck introduced the idea of an intrinsic angular
momentum of $\frac{1}{2}\hbar$ for the electron, finding not only
a definite explanation for the \emph{anomalous} Zeeman effect, but
also establishing since then a connection between spin and the
Exclusion Principle, a connection whose depth they could not
guess.

Pauli's Exclusion Principle remains as a postulate, for Pauli's
own dissatisfaction, as he expressed in his Nobel prize acceptance
lecture in 1946:

\begin{quote}
\emph{``Already in my original paper I stressed the circumstance
that I was unable to give a logical reason for the exclusion
principle or to deduce it from more general assumptions. I had
always the feeling, and I still have it today, that this is a
deficiency." }\cite{pauli-lecture}
\end{quote}
In any case, as inexplicable as it may be, Pauli's Exclusion
Principle seems to beg for a generalization. In fact, it was soon
realized that other particles apart from electrons suffer from the
same inability to share a common quantum state (e.g., protons).
More surprising was the indication that some particles seem to
obey to the exactly opposite effect, being
--- under certain circumstances --- forced to share a common state, as
for instance photons in the stimulated emission phenomenon, thus
calling for a much more drastic generalization of Pauli's
Principle, as we shall see.

\section{Identity and Indistinguishability}

We saw that Pauli's Exclusion Principle intervenes in a wide range
of phenomena, from the chemical bond in the salt on our table to
the formation of stars in distant galaxies. This is because it
applies to electrons and we consider all electrons in the universe
to be \emph{identical}, as well as any other kind of quantum
particles:\\

\textbf{Identical particles} --- \emph{Two particles are said to
be \emph{identical} if all their intrinsic properties (e.g., mass,
electrical charge, spin, colour, ...) are exactly the same.}\\
\\
Thus, not only all electrons are identical, but also all
positrons, photons, protons, neutrons, up quarks, muon neutrinos,
hydrogen atoms, etc. They each have the same defining properties
and behave the same way under the interactions associated with
those properties. This brings us to yet another purely quantum
effect, that of \emph{indistinguishable} particles.

Imagine we have two completely identical classical objects, that
we cannot differentiate in any way. Should we give them arbitrary
labels, we could always --- at least in principle --- keep track
of which object is which by following their respective
\emph{trajectories}. But we know that in quantum mechanics we must
abandon this classical concept. The best information we can get
about some particle's location without measuring it (and thus
disturbing it) is that is has a certain probability of being in a
particular position in space, at a given moment in time. This
information is contained in the particle's \emph{spatial state},
for instance given by:
\begin{equation}
|\psi \rangle_V = \int  \psi(\textbf{r},t) |\textbf{r} \rangle
d^3r,
\end{equation}
where the vectors $|\textbf{r} \rangle$, each representing the
state corresponding to a particular position of the particle in
the three-dimensional Euclidean space, constitute an orthonormal
(continuous) basis of the Hilbert space associated to the position
degree of freedom. The coefficient $\psi(\textbf{r},t)$, also
known as the particle's \emph{wave function}, contains the
probabilistic information about the location of the particle, the
only information available to us prior to a measurement. The
probability of finding the particle in position $\textbf{r}_0$ at
a time $t$ is given by:
\begin{equation}
P(\textbf{r}_0,t)=| \langle \textbf{r}_0 |\psi \rangle_V |^2 =
|\psi(\textbf{r}_0,t)|^2.
\end{equation}
Note that there is usually a volume V outside which
$\psi(\textbf{r},t)$ quickly falls off to zero asymptotically. We
associate the \emph{spread} of the wave function to this volume
$V$, which can evolve in time. Finally, recall that because of the
Heisenberg's uncertainty relations we cannot simultaneously
measure the particle's position and its momentum with an arbitrary
precision (see, for instance, \cite{cohen}).

\begin{figure}[ht]
\begin{center}
\epsfig{file=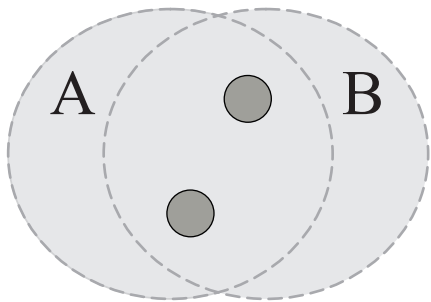}
\end{center}
\caption{This figure represents two identical particles in a
situation where their respective wave functions overlap. It is no
longer unambiguous which region of space each particle occupies.
The particles are indistinguishable --- a purely quantum effect
--- and must now obey the Symmetrization Postulate: there are symmetry
restrictions to the states describing the joint system.}
\label{Fig. PS-Indistinguishable}
\end{figure}

How can we then distinguish identical particles? Their possibly
different internal states are not a good criterium, as the
dynamics can in general affect the internal degrees of freedom of
the particles. The same is valid for their momentum or other
dynamical variables. But their spatial location can actually be
used to distinguish them, as shown in Fig.\ \ref{Fig.
PS-Distinguishable}. Let us imagine we have two identical
particles, one in Alice's possession and the other with Bob. If
these two parties are kept distant enough so that the wave
functions of the particles practically never overlap (during the
time we consider this system), then it is possible to keep track
of the particles just by keeping track of the classical parties.
This situation is not uncommon in quantum mechanics. If, on the
other hand, the wave functions do overlap at some point, then we
no longer know which particle is with which party, as shown in
Fig.\ \ref{Fig. PS-Indistinguishable}. And if we just do not or
cannot involve these classical parties at all, then it is in
general also impossible to keep track of identical particles. In
both these cases, the particles become completely
indistinguishable, they are identified by completely arbitrary
labels, with no physical meaning (as opposed to \emph{Alice} and
\emph{Bob}). In these situations arise very interesting new
phenomena.


\newpage
\vspace*{1.5cm}

\begin{figure}[ht]
\begin{center}
\epsfig{file=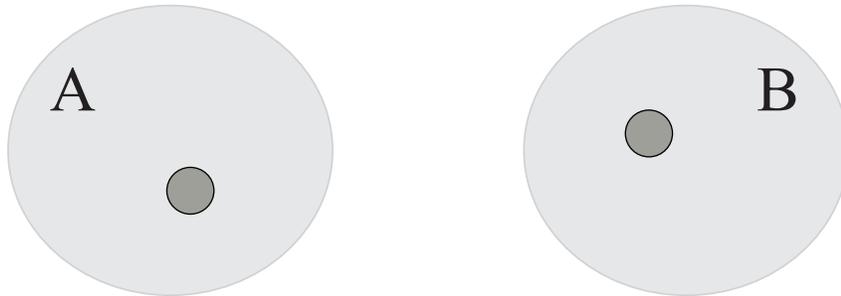}
\end{center}
\caption{This figure represents two distant identical particles,
as well as the spread of their respective wave functions. Each one
of them occupies a distinct region of space, arbitrarily labelled
$A$ and $B$, thus allowing us to distinguish these identical
particles, just as in a classical case.} \label{Fig.
PS-Distinguishable}
\end{figure}

\vspace{1.5cm}

\section{Symmetries, Fermions and Bosons}
\label{Sec. Symmetries}

Let us consider the following example. Imagine that we have two
electrons in an indistinguishable situation, e.g.\ as the one
described in Fig.\ \ref{Fig. PS-Indistinguishable}. We can
arbitrarily label them $1$ and $2$. We also know that one of the
particles has spin up along the $z$ direction and the other has
spin down along the same direction. How shall we describe the
(spin) state of our system? One possibility would be to consider
the vector:
\begin{equation}
| \mu \rangle = | \!\!  \uparrow \rangle _1 | \!\! \downarrow
\rangle _2,
\end{equation}
where $|\!\! \uparrow \rangle_i$ and $|\!\! \downarrow \rangle_i$
represent the two opposite spin components along $z$ for each
particle $i$, and $\{ | \!\! \uparrow \rangle_i, | \!\! \downarrow
\rangle_i \}$ constitutes an orthonormal basis of the
2-dimensional Hilbert space. But, since the particles are
indistinguishable, we could permute their labels and the state of
the system could equally be described by the vector:
\begin{equation}
| \nu \rangle = | \!\! \downarrow \rangle _1 | \!\! \uparrow
\rangle _2.
\end{equation}
Note that the state of the system is the same, but we have two
different vectors that can validly describe it. In fact, taking
into account the superposition principle, a linear combination of
$| \mu \rangle$ and $| \nu \rangle$ will also be a possible
description of our state:
\begin{equation}
\label{Eq. Example kappa}
| \kappa \rangle = \alpha | \!\!
\uparrow \rangle _1 | \!\! \downarrow \rangle _2 + \beta | \!\!
\downarrow \rangle _1 | \!\! \uparrow \rangle _2,
\end{equation}
where $\alpha,\beta \in \cal C$ are chosen such that
$|\alpha|^{2}+|\beta|^{2}=1$. So, we actually have an infinity of
different mathematical descriptions for the same physical state.
This is a consequence of the indistinguishability of particles and
is know as \emph{exchange degeneracy}. How can we then decide
which of the above vectors is the correct description of our
state, i.e.\ which one will allow us to make correct predictions
about measurements or the evolution of the system? Should it be
$|\kappa \rangle$, for which particular values of $\alpha$ and
$\beta$? Also, note that our example could be generalized to more
and other species of particles: the exchange degeneracy appears
whenever we deal with indistinguishable particles. The problem of
finding the correct and unambiguous description for such systems
is thus very general and requires the introduction of a new
postulate for quantum mechanics: the Symmetrization Postulate.\\

\textbf{Symmetrization Postulate} \label{Post. Symmetrization} ---
\emph{In a system containing indistinguishable particles, the only
possible states of the system are the ones described by vectors
that are, with respect to permutations of the labels of those
particles:}
\begin{itemize}

\item \textit{either \emph{completely symmetrical} --- in which
case the particles are called \emph{bosons};}

\item \textit{either \emph{completely antisymmetrical} --- in
which case the particles are called \emph{fermions}.}

\end{itemize}

It is this information that will allow us to lift the exchange
degeneracy. But first, let us consider a general example to
discuss the concepts and issues introduced by this postulate.

Imagine we have a system $\Sigma$ of $N$ indistinguishable
particles, to which we associate the Hilbert space $\HH_N \equiv
\HH^{\otimes N}$. We arbitrarily label the particles with numbers
from $1$ to $N$. The state of the system can be described by:
\begin{equation}
| \Omega \rangle_N \equiv | \alpha \rangle_1 \otimes | \beta
\rangle_2 \otimes \ldots \otimes | \xi \rangle_i \otimes \ldots
\otimes | \tau \rangle_N,
\end{equation}
where $| \xi \rangle_i$ arbitrarily indicates that particle $i$ is
in state $| \xi \rangle \in \HH$. But the above postulate imposes
some symmetries. Let us first define the following terms.

A \emph{completely symmetric} vector $|\psi_S \rangle$ is a vector
that remains invariant under permutations of its $N$ labels (or
components), such that:
\begin{equation}
\hat{P}_N |\psi_S \rangle = |\psi_S \rangle
\end{equation}
for any permutation $P_N$ of those $N$ different labels. In $|
\Omega \rangle_N$, and in most of our other examples, these will
just be the integers from 1 to $N$. Also, note that there are $N!$
such permutations.

Similarly, a \emph{completely antisymmetric} vector $|\psi_A
\rangle$ is a vector that satisfies:
\begin{equation}
\hat{P}_N |\psi_A \rangle = \varepsilon_{P_N} |\psi_A \rangle
\end{equation}
for any permutation $P_N$, and where:
\begin{eqnarray}
\varepsilon_{P_N} = \left\{ \begin{array}{c}
+1 \textrm{ if } P_N \textrm{ is an even permutation}\\
-1 \textrm{ if } P_N \textrm{ is an odd permutation}\\
\end{array} \right. .
\end{eqnarray}

To impose these symmetries to a vector, we can define the
\emph{symmetrizer} and \emph{antisymmetrizer} operators,
respectively given by:
\begin{equation}
\label{Eq. Symmetrizer} \hat{S} \equiv \frac{1}{N!} \sum_{P_N}
\hat{P}_N
\end{equation}
and
\begin{equation}
\label{Eq. Antisymmetrizer}
\hat{A} \equiv \frac{1}{N!} \sum_{P_N}
\varepsilon_{P_N} \hat{P}_N,
\end{equation}
and where the sums are taken over the $N!$ possible permutations
$P_N$ of a set of $N$ elements. We can then apply these operators
to $| \Omega \rangle_N$ to obtain a completely symmetric vector:
\begin{equation}
| \Omega_S \rangle_N \equiv \hat{S} | \Omega \rangle_N,
\end{equation}
as well as a completely antisymmetric vector:
\begin{equation}
| \Omega_A \rangle_N \equiv \hat{A} | \Omega \rangle_N,
\end{equation}
one of which will be the correct description of our system
$\Sigma$, in accordance to the Symmetrization Postulate. Note that
$\hat{S}$ is the projection operator onto the completely symmetric
subspace of $\HH_N$, i.e.\ the Hilbert space spanned by all the
independent completely symmetric vectors of $\HH_N$, that we shall
call $\HH_S$. Analogously, $\hat{A}$ is the projector onto the
completely antisymmetric subspace $\HH_A$. In general, we have the
relation:
\begin{equation}
\HH_N = \HH_S \oplus \HH_A \oplus \HH_m,
\end{equation}
where $\HH_m$ is a subspace of $\HH_N$ with mixed symmetry. We
thus see that, for systems of indistinguishable particles, the
Symmetrization Postulate restricts the Hilbert space of the
system. The only acceptable vectors to describe the system must
lie in either the completely symmetric or in the completely
antisymmetric subspaces. But how do we know which one of these two
exclusive possibilities to choose? In particular, coming back to
the example system $\Sigma$, which of the vectors $| \Omega_S
\rangle_N$ and $| \Omega_A \rangle_N$ represents the state of our
system? This is something that depends on the nature of the
particles, if they are either bosons or fermions respectively. And
to which of these two classes a particle belongs is something that
ultimately can only be determined experimentally.

\subsection{Fermions}
\label{Sec. Fermions}

We call \emph{fermions} identical particles that, in an
indistinguishability situation, can only be found in antisymmetric
states, i.e.\ described by vectors in the system's antisymmetric
subspace. Such states can be constructed using  the
antisymmetrization operator $\hat{A}$ defined in equation
(\ref{Eq. Antisymmetrizer}). It is interesting to note that the
sum of permutations with alternating sign can formally be obtained
using a determinant. This is known as the \emph{Slater
determinant} and offers us a practical way to construct completely
antisymmetric vectors. For instance, $| \Omega_A \rangle_N$ of our
example can be calculated the following way:

\begin{eqnarray}
| \Omega_A \rangle_N = \hat{A} | \Omega \rangle_N  = \frac{1}{N!}
\left|
\begin{array}{cccccc}
| \alpha \rangle_1 & | \beta \rangle_1 & \cdots & | \xi \rangle_1
& \cdots & | \tau \rangle_1 \\
| \alpha \rangle_2 & | \beta \rangle_2 & \cdots & | \xi \rangle_2
& \cdots & | \tau \rangle_2 \\
\vdots & \vdots & \ddots & \vdots
& \ddots & \vdots \\
| \alpha \rangle_i & | \beta \rangle_i & \cdots & | \xi \rangle_i
& \cdots & | \tau \rangle_i \\
\vdots & \vdots & \ddots & \vdots
& \ddots & \vdots \\
| \alpha \rangle_N & | \beta \rangle_N & \cdots & | \xi \rangle_N
& \cdots & | \tau \rangle_N \\
\end{array} \right| .
\end{eqnarray}\\
Note that if there are two particles of the system in the same
state, say $| \alpha \rangle = | \beta \rangle$, then two columns
of the Slater determinant will be equal and the determinant will
be zero. This means that there are no vectors to describe systems
of fermions where two particles are in the same state. This is
just as the Exclusion Principle that we introduced for electrons
in section \ref{Sec. Pauli}, but now for a much broader set of
particles. We conclude that the Symmetrization Postulate not only
includes, but actually generalizes Pauli's Exclusion Principle to
other particles than electrons. Moreover, we can now say that
electrons are fermions. This information allows us to apply the
postulate to our example of the two electrons with anti-aligned
spins discussed in the beginning of section \ref{Sec. Symmetries}
and thus finally lift the respective exchange degeneracy. Let us
assume, for the sake of the argument, that in this example the
total spatial wave function of the particles is symmetrical. The
correct description of this system is then given by the vector in
equation (\ref{Eq. Example kappa}) with the following choice of
coefficients to make it, together with the full state, completely
antisymmetric:
\begin{equation}
\label{Eq. Singlet12}
| \kappa_A \rangle \equiv \frac{1}{\sqrt{2}}
\left( | \!\! \uparrow \rangle _1 | \!\! \downarrow \rangle _2 - |
\!\! \downarrow \rangle _1 | \!\! \uparrow \rangle _2 \right).
\end{equation}
We see that, even for electrons, the Symmetrization Postulate
gives us more information than the Exclusion Principle.

We now have a more physical and operational definition of
fermions: particles that, when indistinguishable, can never be in
the same state. This restriction has very clear consequences when
we study the statistical properties of quantum systems of many
identical particles. All fermions follow the same distribution for
the average number of particles in a certain quantum state, say
$\ell$, in function in terms of the parameters of the system:
\begin{equation}
\label{Eq. Fermi-Dirac}
\langle n_\ell \rangle
=\frac{1}{e^{\frac{E_\ell-\mu}{k_B T}}+1},
\end{equation}
where $\mu$ is the chemical potential per particle, $E_\ell$ is
the energy of the particle in state $\ell$, $T$ is the temperature
of the system and $k_B \simeq 1.38 \times 10^{-23} J K^{-1}$ is
Boltzmann's constant. This is known as \emph{Fermi-Dirac's
statistical distribution} and is a direct consequence of the
antisymmetry of the fermionic states (see, for example,
\cite{landau} for the proof). Note that we always have $\langle
n_\ell \rangle \leq 1$, as we would expect. This distribution was
first calculated for electrons by Enrico Fermi \cite{fermi} in
1926 and its more general relations with quantum mechanics where
established soon after by Paul Dirac \cite{dirac}. This
distribution plays a central in quantum statistics. It is, for
example, fundamental to describe the electronic structure of
solids and their electrical and thermal properties. Note also that
in the limit where particles become distinguishable (e.g., because
of a lower density, a larger separation between energy levels,
etc.) as thermal fluctuations become more important than the
quantum ones, we recover the classical Boltzmann distribution.

We thus have two ways to decide whether a given type of particles
are fermions: they explicitly obey to the Exclusion Principle
(more easily observable in systems with few particles), and they
follow the Fermi-Dirac distribution (a criterion more adequate for
systems of many particles). In any case, an experimental proof is
necessary. It has also been observed that all fermions have
half-integer spin. It should be noted that this surprising and
useful property is not part of the definition of fermions, or at
least does not need to be. The connection between spin and
statistics will be further discussed in section \ref{Sec.
Spin-Statistics}. Particles that are nowadays known to be fermions
include: electrons and in fact all leptons, quarks, protons,
neutrons, baryons in general, $^3$He, etc.

Finally, note that the symmetry requirements of the Symmetrization
Postulate are, in the second quantization formalism\footnote{The
reader unfamiliar with this formalism is referred to appendix A
for a brief introduction.}, replaced by imposing certain algebraic
relations on the fermionic creation and annihilation operators,
respectively denoted $\aa_{i}^{\dag}$ and $\aa_{i}$. These are
defined by:
\begin{equation}
\left\{
\begin{array}{l}
\aa_{i}^{\dag} |\,n_1, \ldots, n_{i-1}, 0_i, n_{i+1}, \ldots
\rangle = (-1)^{p_i} |\,n_1, \ldots, n_{i-1}, 1_i, n_{i+1}, \ldots
\rangle \\
\aa_{i}^{\dag} |\,n_1, \ldots, n_{i-1}, 1_i, n_{i+1}, \ldots
\rangle = 0
\end{array}
\right.
\end{equation}
and
\begin{equation}
\left\{
\begin{array}{l}
\aa_{i} |\,n_1, \ldots, n_{i-1}, 1_i, n_{i+1}, \ldots \rangle =
(-1)^{p_i} |\,n_1, \ldots, n_{i-1}, 0_i, n_{i+1}, \ldots
\rangle \\
\aa_{i} |\,n_1, \ldots, n_{i-1}, 0_i, n_{i+1}, \ldots \rangle = 0
\end{array}
\right. , \vspace{0.05cm}
\end{equation}
with $p_i=\sum_{k<i} N_k$, where $N_k$ is the eigenvalue of the
number operator $\hat{N}_k \equiv \aa^\dag_k \aa_k$. The condition
of complete antisymmetry for vectors describing indistinguishable
fermions is then replaced by the following anti-commutation
relation on the creation and annihilation operators describing the
system:

\begin{equation}
\label{Eq. Anti-commutation}
\left[\aa_{i},\aa_{j}^{\dag}\right]_{+} \equiv
\aa_{i}\aa_{j}^{\dag} +
\aa_{j}^{\dag}\aa_{i}=\delta_{ij}\,\hat{I}, \vspace{0.2cm}
\end{equation}
where $i$ and $j$ are sets of labels for (in general) different
modes, $\delta_{ii}= 1$ and $\delta_{ij}= 0$ for $i \neq j$. The
two formalisms are equivalent.

\subsection{Bosons}
\label{Sec. Bosons}

We call \emph{bosons} identical particles that, in an
indistinguishability situation, can only be found in symmetric
states, i.e.\ described by vectors in the system's symmetric
subspace. Such states can be constructed using  the symmetrization
operator $\hat{S}$ defined in equation (\ref{Eq. Symmetrizer}).
The average number of bosons in a certain quantum state, say
$\ell$, is given by the \emph{Bose-Einstein statistical
distribution} \cite{landau}:
\begin{equation}
\langle n_\ell \rangle =\frac{1}{e^{\frac{E_\ell-\mu}{k_B T}}-1},
\end{equation}
where $\mu$ is the chemical potential per particle and $\mu < 0$,
$E_\ell$ is the energy of the particle in state $\ell$, $T$ is the
temperature of the system and $k_B$ is Boltzmann's constant.
Formally, this distribution differs only by a sign when compared
with Fermi-Dirac's one, given by equation (\ref{Eq. Fermi-Dirac}),
but the differences are actually great: in this case, $\langle
n_\ell \rangle$ is unbounded. For bosons there is no Exclusion
Principle, on the contrary. Firstly, an arbitrary number of them
can occupy the same state. Secondly, in certain cases the
Symmetrization Postulate actually forces the particles to share a
common state, other alternatives being incompatible with the
symmetry requirements. It is almost like bosons follow some kind
of \emph{aggregation principle}, opposing Pauli's one.

The Bose-Einstein distribution was first proposed by Satyendranath
Bose in 1924 for a gas of photons \cite{snbose} and very soon
after generalized by Albert Einstein to an ideal monoatomic gas
\cite{einstein1, einstein2}. Einstein noticed then that, under a
certain critical temperature, a fraction of the particles would
gather in the energy ground state \cite{einstein2}; this fraction
grows as the temperature decreases and eventually includes all the
particles when the temperature is zero. This phase transition is
known as the \emph{Bose-Einstein condensation} and has only been
directly observed (with non-interacting particles) in 1995 ---
seventy years after Einstein predicted it --- using Rubidium atoms
at $20\: nK$ \cite{BEC}. As in the case of fermions, the
Bose-Einstein distribution tends to Boltzmann's as particles
approach the classical regime.

For bosons there is also a connection between spin and statistics.
In fact, we believe that all particles of integer spin (including
0) are bosons: photons, gluons, $W^{\pm}$, $Z^0$, mesons, $^4$He
and any other atoms or nuclei with integer spin, etc. This
property will be discussed in more detail in section \ref{Sec.
Spin-Statistics}.

Finally, a word about the treatment of bosons in second
quantization. In this formalism we define the bosonic field
creation operator $\aa_{i}^{\dag}$ by:
\begin{equation}
\aa_{i}^{\dag} |\,n_1, \ldots, n_{i-1}, n_i, n_{i+1}, \ldots
\rangle = \sqrt{n_i+1} |\,n_1, \ldots, n_{i-1}, n_i+1, n_{i+1},
\ldots \rangle,
\end{equation}
and the bosonic field annihilation operator $\aa_{i}$ by:
\begin{equation}
\aa_{i} |\,n_1, \ldots, n_{i-1}, n_i, n_{i+1}, \ldots \rangle =
\sqrt{n_i} |\,n_1, \ldots, n_{i-1}, n_i-1, n_{i+1}, \ldots \rangle
.\end{equation}
Note that we use exactly the same notation for the bosonic and
fermionic field operators (defined in section \ref{Sec.
Fermions}): although this can at first seem a source of confusion,
this notation will be useful later. But, of course, one should
keep in mind at all times that these are different operators by
definition, with different commutation properties. In particular,
the requirement of complete symmetry for bosons is obtained
imposing the following commutation condition on the creation and
annihilation operators describing the system:

\begin{equation}
\label{Eq. Commutation} \left[\aa_{i},\aa_{j}^{\dag}\right]_{-}
\equiv \aa_{i}\aa_{j}^{\dag} - \aa_{j}^{\dag}\aa_{i}=\delta_{ij}
\,\hat{I}, \vspace{0.2cm}
\end{equation}
where $i$ and $j$ are sets of labels for (in general) different
modes, $\delta_{ii}= 1$ and $\delta_{ij}= 0$ for $i \neq j$.

\subsection{The Spin-Statistics Connection}
\label{Sec. Spin-Statistics}

We saw that to determine whether a given particle is a fermion or
a boson, we need to investigate its statistical behaviour in the
presence of (at least one) other identical particles, when they
are all indistinguishable. Of course, we can expect that if a
composite particle is made of bosons, or of an even number of
fermions, then it should be a boson. And if, on the other hand,
the particle is composed of an odd number of fermions, then it
should be a fermion itself. But for the particles we believe to be
fundamental a direct study of their statistical nature may be
required. From the experimental point of view, such a study can
represent quite a difficult challenge. For instance, we may be
dealing with rare particles, difficult to observe (e.g.,
neutrinos), to find in free form (e.g., quarks and gluons) or even
to produce (e.g., the \emph{top} quark), or particles that are
just too unstable or short-lived (e.g., the $\tau$ lepton) or yet
to be observed (e.g., the graviton). Indirect methods could also
help us reach a conclusion, but before any of that a simple and
intriguing property can actually come to our rescue: the
\emph{spin-statistics connection}.\\

\textbf{Spin-Statistics Theorem} --- \textit{Particles with
integer spin are bosons. Particles with half-integer spin are
fermions.}\\

This is not only a widely known empirical rule in Physics, but in
fact a theorem, even if its proofs are not all completely clear
and free from controversy. Thanks to it, it is very easy to
determine whether some particle is either a fermion or a boson. In
particular, this criterion works also for composite particles and
is consistent with the previous conclusion that particles composed
of an arbitrary number of bosons and/or of an even number of
fermions are bosons and that particles composed of an odd number
of fermions are fermions. It is quite surprising to find such a
connection between the spin of a particle and its statistical
nature, a connection whose origins I think are still not well
understood.

The first and reference proof of the Spin-Statistics Theorem is
usually and fairly attributed to a 1940 article by Pauli
\cite{pauli-theorem}, despite some earlier contributions towards
this problem (see \cite{sudarshan} for a historical and technical
account of those works). Pauli's proof, referring only to free
particles, is based on two (reasonable) assumptions:
\begin{itemize}

\item By assuming that the energy must be a positive quantity, he
concludes that particles with half-integer spins cannot be bosons;

\item By assuming \emph{microcausality}\footnote{The usual special
relativity restriction that the measurement of a physical system
cannot influence another if the two are space-like separated.}, he
concludes that particles with integer spin cannot be fermions.

\end{itemize}
Markus Fierz had just proven a year earlier that, under these same
assumptions, particles with integer spin could be bosons and
particles with half-integer spin could be fermions \cite{fierz},
and with those results Pauli could conclude his proof. Pauli
wanted to present the spin-statistics connection as a direct
consequence of Special Relativity, but his \emph{negative} and
\emph{asymmetric} proof depended also on the condition of positive
energy. Several other proofs appeared over the following decades,
more or less along the lines of Pauli's work and always in the
context of Relativistic Quantum Mechanics, and in particular of
Quantum Field Theory (see \cite{sudarshan} and references
therein). Note also Steven Weinberg's proof \cite{weinberg} based
solely on the Lorentz invariance of the
\emph{S-matrix}\footnote{The \emph{S-matrix formalism} is an
alternative approach to relativistic quantum physics based on the
unitary S-matrix that encodes all the information on all possible
scattering processes. Formally, the S-matrix is the realization of
the isomorphism between the \emph{in} and \emph{out} Fock spaces.
For more details, see for instance \cite{weinberg-book}.},
breaking with some traditions in the approach to the theorem
\cite{michela}. These (and other similar) works were a source of
debates and brought some extensions and clarifications to details
of other proofs, but there was no real advance in our
understanding of this connection. To put it in Richard Feynman's
words:

\begin{quote}
\emph{``It appears to be one of the few places in physics where
there is a rule which can be stated very simply, but for which no
one has found a simple and easy explanation. [...] This probably
means that we do not have a complete understanding of the
fundamental principle involved."} \cite{feynman}
\end{quote}

If this relation between spin and statistics is indeed a
relativistic effect, or at least if the proof of the respective
theorem relies on Special Relativity, then the spin-statistics
connection has to be introduced in (non-relativistic) Quantum
Mechanics as a postulate. Recently, some non-relativistic proofs
have been proposed \cite{balachandran, berry, duck}, but the
claims are not uncontroversial (see \cite{duck}). It remains that
the connection between spin and statistics is an empirical rule,
and for now that is probably how it should be introduced in the
context of Quantum Mechanics.

Research on the connection between spin and statistics continues
also in other directions, in particular in the study of particles
obeying alternatives statistics. These will be briefly discussed
in section \ref{Sec. Alternative}. Note also that in 1974 a new
fundamental symmetry was proposed \cite{wess},
\emph{supersymmetry}, that transforms bosons into fermions and
vice-versa, providing a framework for the unification of gravity
with the other interactions (for more details, see for example
\cite{freund}). To this date, no experimental evidence of
supersymmetric particles has been found though \cite{pdg}. Nor, in
fact, any evidence of violations of the predictions of the
Spin-Statistics Theorem --- maybe our best and clearest proof so
far that it holds.

\subsection{Discussion of the Symmetrization Postulate}

Historically, Werner Heisenberg was the first to show --- in 1926,
in the context of wave mechanics  \cite{heisenberg1, heisenberg2}
--- that the states of a system of identical particles are restricted
to specific symmetry classes which cannot be transformed into each
other. To decide which particles should be associated to a
particular symmetry is something that must ultimately be
determined by observation. The Symmetrization Postulate matches
the study of such symmetries with our empirical knowledge: as far
as we know today, there are two classes of particles in Nature
according to their collective behaviour in indistinguishable
situations. These are, of course, bosons and fermions: no
particles have been found so far that under the same circumstances
could be described by vectors that are neither symmetrical nor
antisymmetrical (see section \ref{Sec. Alternative}). It is
important to note that none of this could have been deduced from
the other postulates of Quantum Mechanics. Yet, the Symmetrization
Postulate is rarely evoked. In textbooks, it is never presented
with the postulates of Quantum Mechanics, but rather postponed for
a final chapter about identical particles, if it appears at all as
such. Pedagogical and epistemological issues aside, such an
approach gives in my opinion an incomplete picture of the theory,
both formally and physically, responsible for some ignorance and
misunderstandings. It is true that the Symmetrization Postulate
applies only to indistinguishable particles and is, in that sense,
less general than the other postulates of Quantum Mechanics.
Moreover, it imposes a limitation on the \emph{Hilbert (or state)
space postulate}, by restricting the state space of the system to
its completely symmetric or antisymmetric subspaces. But, on the
other hand, it is also true that what makes Quantum Mechanics so
interesting is that it can describe matter (and fields) around us,
and for that, as we saw, it is necessary to introduce the Pauli
Exclusion Principle, or --- more generally
--- the Symmetrization Postulate. This is, after all, the way we
formally introduce fermions and bosons in Quantum Mechanics.

The awkward status of the Symmetrization Postulate probably
reflects our discomfort in trying to understand some of these
issues, as Pauli mentioned already in 1946:
\begin{quote}
\emph{``Of course in the beginning I hoped that the new quantum
mechanics, with the help of which it was possible to deduce so
many half-empirical formal rules in use at that time, will also
rigorously deduce the exclusion principle. Instead of it there was
for electrons still an exclusion: not of particular states any
longer, but of whole classes of states, namely the exclusion of
all classes different from the antisymmetrical one. The impression
that the shadow of some incompleteness fell here on the bright
light of success of the new quantum mechanics seems to me
unavoidable." }\cite{pauli-lecture}
\end{quote}
We have a very simple and elegant rule, crucial for the success of
Quantum Mechanics as a theory describing Nature, undisputed and
apparently free from interpretations, unlike other quantum rules.
We have no clear understanding of its origin, not even within the
framework of a more sophisticated theory than Quantum
Mechanics\footnote{As opposed to the spin-statistics connection, a
property that we believe can be derived in the context of
relativistic Quantum Mechanics, as we saw in section \ref{Sec.
Spin-Statistics}.}, and are thus happy (or forced) to accept it as
an empirical rule, but one separated from the other postulates of
the theory, even if it is not less mysterious than these, nor less
fundamental as far as we now today.

Let us now analyse the statement of the Symmetrization Postulate.
The principle applies only to indistinguishable particles. This is
exactly how we have presented it here, although some prefer to
state it as applying to identical particles in general. \emph{A
priori}, the latter approach implies that, for instance, every
electron in the universe would have to be antisymmetrized with
respect to all the others. And every other kind of particle would
have to obey to a similar requirement, adequate to its statistical
nature. This picture is of course very unsatisfactory, but ---
conveniently --- it can be shown that the (anti)symmetrization
terms have a vanishing probability when we can distinguish the
identical particles (see, for example, \cite{cohen}). Thus, the
Symmetrization Postulate ends up being applied only to
indistinguishable particles. Note also that, given a system, we
often consider separate vectors to describe its spatial and its
internal degrees of freedom. Then, the symmetries of each vector
are not independent, as they must consistently contribute to the
symmetry requirement of the full vector describing the state of
the system. This is a natural consequence of the postulate and is
not implied by the concept of \emph{complete symmetry}, which
specifically refers to all possible permutations that can be
considered. Finally, note that if a composite particle is made of
bosons or of an even number of fermions then it is a boson, and if
it is composed of an odd number of fermions then it is itself a
fermion.

All evidence to this date points to the fact that quantum
indistinguishable particles either follow Fermi-Dirac's or
Bose-Einstein's statistics, something that is generally referred
to as \emph{particle} or \emph{quantum statistics}. Accordingly,
the Symmetrization Postulate, with its complete symmetry or
antisymmetry requirement, rules out the consideration of other
statistics from the onset. Yet, without this restriction, Quantum
Mechanics would actually allow unusual symmetries, more complex
than the previous ones. As we shall see in section \ref{Sec.
Alternative}, consistent theories have been developed allowing for
small deviations from the conventional statistics which might have
been undetected in the experiments performed so far.

Finally, I would like to point out the interesting and possibly
deep fact that, according to the Standard Model of Particle
Physics, the fundamental constituents of matter --- quarks and
leptons --- are fermions, whereas the force carriers are bosons.
Note also that, within the framework of Quantum Field Theory, the
equivalent to the Symmetrization Postulate is introduced by
imposing local (equal-time) commutation and anti-commutation rules
on the field operators, similar to the ones given by equations
(\ref{Eq. Commutation}) and (\ref{Eq. Anti-commutation}).

\section{Experimental Evidence and Alternative Statistics}
\label{Sec. Alternative}

Particle statistics, or the existence of fermions and bosons and
their respective properties, is well established, accounting for a
series of different physical phenomena and having been subjected
to extensive experimental corroboration. The origin of the
Symmetrization Postulate itself lies in the attempt to explain
experimental data, in particular the energy spectra of atoms. In
1925, Pauli proposed the Exclusion Principle (as we discussed in
section \ref{Sec. Pauli}). Soon after, Heisenberg was able to
introduce it, together with spin, in the context of wave mechanics
and used the symmetry properties of the electrons' wave functions
(decomposed in spatial and spin wave functions) to explain the two
classes of spectral lines observed for Helium: \emph{parahelium}
and \emph{orthohelium} \cite{heisenberg2}. Here came into play the
\emph{exchange interaction}, relating the energy levels occupied
by the electrons to the symmetry of their wave functions and the
corresponding spatial and spin states, a purely quantum effect
associated to the Pauli Principle.

As we saw, the Symmetrization Postulate generalized the Pauli
Exclusion Principle to more particles than just electrons, i.e.\
fermions, and introduced a fundamentally new type of particles,
i.e.\ bosons. Moreover, it applies to composite systems in
general, where one single particle may be enough to change the
system's statistical nature and corresponding behaviour. A
paradigmatic example is given by the two isotopes of Helium:
$^3He$ (a fermion) and $^4He$ (a boson). The latter exhibits
remarkable properties when cooled down to temperatures below $2\:
K$: it becomes a \emph{superfluid} liquid, with practically no
viscosity and capable of flowing with no friction through tiny
capillaries \cite{superfluidity} \footnote{Note that at even lower
temperatures, below $3\: mK$, $^3He$ can also exhibit a superfluid
phase \cite{osheroff,leggett}, as the fermionic Helium atoms pair
up to form bosonic quasiparticles.}. Particle statistics helps us
explain and model a number of other phenomena, as the ones pointed
out in section \ref{Sec. Pauli}, as well as covalent bonding,
ferromagnetism, superconductivity, Bose-Einstein condensation, the
formation of stars, the structure of hadrons, etc., always in
accordance with experimental observations. Note also some recent
attempts to do quantum information processing based solely on the
use of the statistical properties of particles \cite{omar,
paunkovic, bose}.

Despite of all our understanding of particle statistics, it is
still quite mysterious why (or how) fermions with common values in
their internal degrees of freedom will resist being brought close
together, as in the dramatic example of the formation of neutron
stars, this resistance resulting in an effective force, completely
different from the other interactions we know. This force is
sometimes referred to as \emph{Pauli pressure}. Equally intriguing
is the tendency of photons to bunch together, as in the
\emph{Hanbury Brown and Twiss effect} \cite{HBT}, in which case
the interaction seems to have exactly the opposite effect.

\begin{figure}[ht]
\begin{center}
\epsfig{file=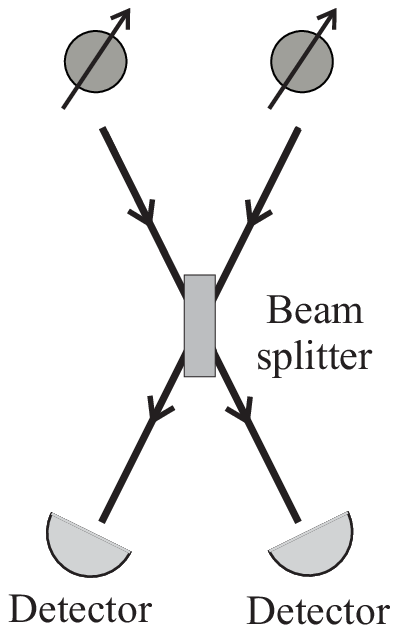, width=2in}
\end{center}
\caption{This figure represents two identical fermions
antibunching out of a $50/50$ beam splitter. If the two particles
are initially in the same internal state --- e.g., two electrons
with aligned spins, as represented by the arrows --- then the
Symmetrization Postulate (or just Pauli's Exclusion Principle)
imposes that they always leave the beam splitter in separate arms.
This phenomenon has recently been observed experimentally for
electrons \cite{yamamoto}.} \label{Fig. PS-Antibunching}
\end{figure}

One interpretation for these effects lies in interference. This
can best be observed using a beam splitter. Here we use the term
\emph{beam splitter} in a generic sense, referring not only to the
common optical element (a partially silvered mirror) used with
photons, but also to any device presenting an analog behaviour for
other kinds of particles, as it was recently proposed for
electrons, using a quantum dot and nanowires to direct the
incoming and outgoing particles \cite{yamamoto}. Note that two
photons with the same frequency and polarization impinging
simultaneously in a $50/50$ beam splitter will always \emph{bunch}
\cite{mandel}, i.e.\ they will always come out together in the
same output arm, as it is shown in Fig. \ref{Fig. PS-Bunching}.
Equivalently, two electrons with the same spin projection
impinging simultaneously in a $50/50$ beam splitter will always
\emph{antibunch} \cite{yamamoto}, i.e.\ they will always come out
separately, one in each output arm, as illustrated in Fig.
\ref{Fig. PS-Antibunching}. When the particles meet in the beam
splitter under the circumstances described above they are fully
indistinguishable and the Symmetrization Postulate must then be
taken into account. The bunching and antibunching of these
particles are effects of particle statistics. Here, they can be
interpreted as the result of constructive and destructive
interference. Note that these results can easily be generalized:
in such balanced beam splitters two indistinguishable particles
will always bunch if they are bosons, and always antibunch if they
are fermions. Another very interesting aspect is that performing
simple path measurements on the output particles can offer us
(probabilistic) information about their internal states.\\


\begin{figure}[ht]
\begin{center}
\epsfig{file=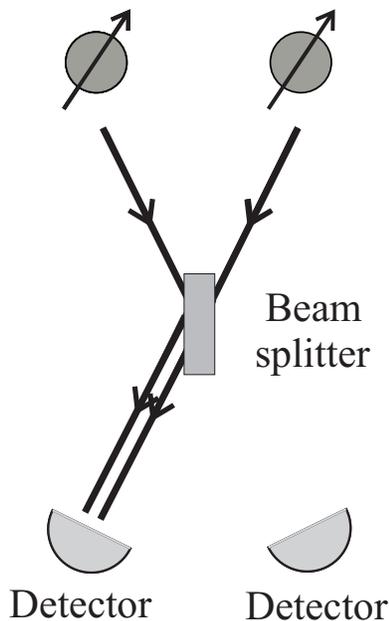, width=2in}
\end{center}
\caption{This figure represents two identical bosons bunching to
the left out of a $50/50$ beam splitter. If the two particles are
initially in the same internal state --- e.g., two photons with
the same polarization, as represented by the arrows --- then the
Symmetrization Postulate imposes that they always leave the beam
splitter in the same arm, either to the left or to the right, each
possibility being equally likely. This phenomenon has been
observed experimentally for photons \cite{mandel}.} \label{Fig.
PS-Bunching}
\end{figure}

Particle statistics, and in particular the Symmetrization
Postulate, represent a limitation on Quantum Mechanics. There is
nothing, apart from our empirical knowledge, that would lead us to
impose such constraints. From the theoretical point of view, it is
perfectly possible to consider different constraints, i.e.\ to
have quantum mechanical theories following \emph{alternative
statistics}, more general than the one we know. Moreover, it is
also possible that such alternative constraints haven been masked
by experiments performed so far.

Alternative statistics have actually been studied. For instance,
\emph{parastatistics} \cite{green} proposes the following
generalization of fermions and bosons:
\begin{itemize}

\item \emph{parafermions} of order $n$, for which a maximum of $n$
indistinguishable such particles can be in a symmetric state,
i.e.\ at most $n$ particles can occupy the same state;

\item \emph{parabosons} of order $n$, for which a maximum of $n$
indistinguishable such particles can be in an antisymmetric state.

\end{itemize}
Note that the case $n=1$ corresponds to the usual particle
statistics and that, for all known particles, parastatistics is
clearly violated for any larger $n$. Yet, in the early sixties,
when the quark model was first proposed, it seemed that baryons
(particles composed of three quarks, such as the proton) could
only be described by a symmetric wave functions, even though they
were fermions. At the time it was considered that quarks could
obey parastatistics \cite{greenberg}, but the problem was later
solved with the introduction of the \textit{colour} internal
degree of freedom in quarks \cite{nambu} --- the charge of the
strong interaction --- which implied an antisymmetric wave
function to describe baryons, as one would expect according to the
Symmetrization Postulate, and thus confirming particle statistics
in the end.

Another alternative to particle statistics are \emph{quons}
\cite{greenberg-quons}, a continuous interpolation between
fermions and bosons. In such a theory, the field operators
describing the states of the particles obey the following
relation:
\begin{equation}
\aa_{i}\aa_{j}^{\dag} - q
\aa_{j}^{\dag}\aa_{i}=\delta_{ij}\hat{I},
\end{equation}
where $q \in [-1,1]$. The limits $q=-1$ and $q=1$ correspond to
fermions and bosons respectively, where we recover the
anticommutation and commutation relations given by equations
(\ref{Eq. Anti-commutation}) and (\ref{Eq. Commutation}). Quons
offer us a setting to extend theses relations infinitesimally (and
continuously) and thus model and investigate small deviations to
Fermi-Dirac and Bose-Einstein statistics. No such violations have
been found up to this date. In fact, several high precision
experiments have been performed recently looking for direct
evidence of the violation of particle statistics or the
spin-statistics connection and have not only confirmed that they
hold --- as one would expect
--- but also established very low probability bounds for
a violation to occur \cite{demille, deilamian, javorsek, tino,
modugno, mazzotti}.

Finally, let us mention \emph{anyons}, quasiparticles in two
dimensions that obey fractional statistics, i.e.\ the permutation
of two of them can give \emph{any} phase \cite{wilczek}, not only
the usual $+1$ and $-1$ associated to bosons and fermions. For
example, the set composed of a charged particle orbiting around a
magnetic flux tube has this property. Anyons are believed to play
a role in the fractional quantum Hall effect \cite{halperin}, and
possibly also in high temperature superconductivity
\cite{laughlin}. Recently, it was proposed to use anyons for
quantum computation \cite{kitaev}, exploring the intrinsic
fault-tolerant properties that these objects can offer.

\section{Summary and Concluding Remarks}

Quantum Mechanics, a theory at the basis of current fundamental
Physics, gives us an incredibly accurate description of Nature at
the atomic scale. Yet, its descriptive power would be very limited
without the introduction of the Symmetrization Postulate. This is
a very simple rule that encodes in an elegant way the observed
fact that all known quantum particles belong to one of two
possible classes given their behaviour in indistinguishable
situations: they are either \textit{fermions} or \textit{bosons}.
We have no clear understanding of the origin of this behaviour,
not even within the framework of a more sophisticated theory, and
are thus happy (or forced) to accept it as an empirical rule. Yet,
the Symmetrization Postulate is often neglected (if not ignored)
when compared to the other postulates of the theory, even if it is
not less mysterious than these, nor less fundamental as far as we
now today. This unfair attitude has hindered the research on the
origin of quantum statistics, as well as on the understanding of
the origin of the Spin-Statistics connection, still today one of
the greatest mysteries of theoretical Physics.

\section*{Acknowledgements}

I would like to thank L. Hardy, P. Knight, J. Jones, V. Vedral and
V. Vieira for their valuable comments about this article, as well
as M. Massimi and G. Milhano for some useful remarks. I would also
like to thank Funda\c{c}\~{a}o para a Ci\^{e}ncia e a Tecnologia
(Portugal) and the 3rd Community Support Framework of the European
Social Fund for financial support under grant SFRH/BPD/9472/2002,
and FCT and EU FEDER through project POCI/MAT/55796/2004 QuantLog.\\

\section*{Appendix A -- Second Quantization Formalism}

The formalism used throughout most of this article, sometimes
referred to as \textit{first quantization}, is standard to
describe non-relativistic Quantum Mechanics, but the
\textit{second quantization} formalism, introduced in this
appendix, could have been used just as well. The second-quantized
theory is more general though: it contains non-relativistic
Quantum Mechanics, but it is a full-fledged relativistic theory
that can describe new processes, such as the creation and
annihilation of particle-antiparticle pairs.

The second quantization formalism is used in Quantum Field Theory,
where different types of particles are described as quanta of
different fields. For instance, a photon is a quantum of the
electromagnetic field. Different photons, i.e., photons with
different energies and polarization, correspond to different modes
of the field. To each of these modes, i.e, to each possible state
of the photons, we associate a \emph{number state} that keeps
track of the number of particles/quanta in that state/mode.
Imagine the following restricted situation, where we have only:
\begin{itemize}
\item 2 photons with energy $E_1$ and polarization $H$
(horizontal);

\item 0 photons with energy $E_1$ and polarization $V$ (vertical);

\item 3 photons with energy $E_2$ and polarization $H$;

\item 2 photons with energy $E_2$ and polarization $V$.
\end{itemize}
Then, we could describe it by the following number state:
\begin{equation}
| 2,0,3,2 \rangle = |2 \rangle _{E_{1},H}|0 \rangle _{E_{1},V}|3
\rangle _{E_{2},H}|2 \rangle _{E_{2},V}.
\end{equation}

Formally, we use the \emph{creation operator} $\hat{a}^\dag_m$ and
the \emph{annihilation operator} $\hat{a}_m$ to respectively
increase or decrease by one the \emph{population} of mode $m$.
Note that $\aa^\dag_m | 0 \rangle$ can be also be used as a
description of $| 1 \rangle_m$, with $| 0\rangle$ representing the
\emph{vacuum state}, where all modes are unpopulated. The
population of a mode is given by the eigenvalue of the
\emph{number operator}:
\begin{equation}
\hat{N}_m \equiv \aa^\dag_m \aa_m.
\end{equation}
Note also that:
\begin{equation}
\label{Eq. Vacuum} \aa_m | 0 \rangle = 0.
\end{equation}
This formalism can be applied not only to photons, but to bosons
in general. In fact, it can also be applied to fermions, although
in this case the creation and annihilation operators will be
different, as we can see in sections \ref{Sec. Fermions} and
\ref{Sec. Bosons}.


\end{document}